\documentclass{aa}
\usepackage{epsfig}
\begin{document}
\thesaurus{03(11.02.1; 11.02.2; 11.04.1;12.07.1)}
\title{The extreme high frequency peaked BL Lac 1517+656\thanks{Based on observations from the German-Spanish Astronomical Center, Calar Alto, operated by the Max-Planck-Institut f\"ur Astronomie, Heidelberg, jointly with the Spanish National Commission for Astronomy}}
\author{V. Beckmann\inst{1,}\inst{2}, N. Bade\inst{1}, and O. Wucknitz\inst{1}}
\offprints{vbeckmann@hs.uni-hamburg.de}
\institute{Hamburger Sternwarte, Gojenbergsweg 112, D-21029 Hamburg, Germany 
 \and Osservatorio Astronomico di Brera, Via Brera 28, I-20121 Milano, Italy}
\date{Received date; accepted date}
\titlerunning{The extreme BL Lac 1517+656}
\maketitle
\begin{abstract}
We present optical spectroscopy data that allowed a measurement of the redshift for the X-ray selected BL Lacertae object
\object{1517+656}. With a redshift of $z = 0.702$
this object has an absolute magnitude of $M_{B} = -26.4$ and is also an 
extremely powerful radio and X-ray source. Although being a high frequency 
peaked BL Lac, this object is one of the most luminous BL Lac objects known so 
far. Being also a candidate for gravitational lensing, this object is of high 
interest for the BL Lac research. Assuming several cosmological models and a realistic redshift for the lensed object, we 
find that 1517+656 has a mass $> 2 \cdot 10^{12} M_{\sun}$ and a high velocity 
dispersion $> 350 \; {\rm km \; sec^{-1}}$. 
\keywords{BL Lacertae objects: general - BL Lacertae objects: individual:
  1517+656 - Galaxies: distances and redshifts - Cosmology: gravitational 
  lensing}
\end{abstract}
\section{Introduction}
The physical nature of BL Lacertae objects is not well understood yet. The
most common view about BL Lac objects is that we are looking into a highly
relativistic jet (Blandford \& Rees \cite{blandford}). This model can explain
several observational parameters, but there are still unsolved problems like the nature of the mechanisms that generate and
collimate the jet or the physical nature and evolution along the jet. An important question is also, if there is a fundamental
difference between BL Lac objects that are found because of their emission in the
radio or in the X-ray range respectively. In order to study the nature of this class of BL Lac,  extreme objects help to give constraints on the physics which is involved. One of the greatest problems while studying
BL Lac is the difficulty in determining their redshifts because of the
absence of strong emission and absorption lines. Usually large telescopes and
long exposure times are needed to detect the absorption lines in the
surrounding host galaxy.\\
\section{History of 1517+656}
Even though 1517+656 is an X-ray selected BL Lac, this object was
detected in the radio band before being known as an X-ray source. It was first
noted in the {\em NRAO Green Bank} $4.85 \; {\rm GHz}$ catalog with a radio flux 
density of $39 
\pm 6 \; {\rm mJy}$ (Becker et al. \cite{becker}) and was also included in
the {\em 87 Green Bank Catalog of Radio Sources}
with a similar flux density of $35 \; {\rm mJy}$ (Gregory \& Condon \cite{87GB}) 
but in both cases without identification of the source. 
The NRAO Very Large
Array at $1.4 \; {\rm GHz}$ confirmed 1517+656 as
having an unresolved core with no evidence of extended emission although a
very low surface brightness halo could not be ruled out (Kollgaard et
al. \cite{kollgaard}). 
The source was first included as an X-ray source in the
HEAO-1 A-3 Catalog 
and was also detected in the {\em Einstein} Slew Survey (Elvis et al. 
\cite{elvis}) in the soft X-ray band ($\sim 0.2-3.5 \; {\rm keV}$) the 
Imaging Proportional Counter
(IPC, Gorenstein et al. \cite{IPC}). The IPC count rate was
$0.91 \; {\rm cts \; sec^{-1}}$, but the total Slew Survey exposure time was
only $13.7 \; {\rm sec}$. Even though 1517+656 by then was a confirmed BL Lac
object (Elvis et al. \cite{elvis}) with an apparent magnitude of $B = 15.5 \; {\rm mag}$, no redshift data
were available. Known as a bright BL Lac, 1517+656 has been studied several times
in different wavelengths in the recent years. Brinkmann \& Siebert (\cite{brinkmann}) presented ROSAT PSPC ($0.07-2.4 \; {\rm keV}$) data and
determined the flux to $f_{\rm X} = 2.89 \; \cdot 10^{-11} \; {\rm erg \; 
cm^{-2} \; sec^{-1}}$ and the spectral index to $\Gamma = 2.01 \pm 0.08$
\footnote{The energy index $\alpha_{E}$ is related to the photon index $\Gamma
  = \alpha_{E}+ 1$}. Observations of 1517+656 with {\em BeppoSAX} in the $2 -
  10 \; {\rm keV}$ band in March 1997 gave an X-ray flux of $f_{\rm X} = 
1.03 \; \cdot 10^{-11} \; {\rm erg \; cm^{-2} \; sec^{-1}}$ and a steeper 
spectral slope of $\Gamma = 2.44 \pm 0.09$ 
(Wolter et al. \cite{wolter}). 
The Energetic Gamma Ray Experiment Telescope (EGRET, Kanbach et
al. \cite{EGRET}) on the {\em Compton Gamma
  Ray Observatory} did not detect 1517+656 but gave an upper flux limit of\\ 
$8 \cdot 10^{-8} \; {\rm photons \;
  cm^{-2} \; sec^{-1}}$ for $E > 100 \, {\rm MeV}$ (Fichtel et
al. \cite{fichtel}). 
In the hard X-rays 1517+656 was first detected with OSSE with \mbox{$3.6 \pm
  1.2 \cdot 10^{-3} \; {\rm photons \; cm^{-2} \; sec^{-1}}$} at $0.05 - 10 \;
{\rm MeV}$ (McNaron-Brown et
al. \cite{mcnaron}). The BL Lac was then detected in the EUVE-All-Sky Survey
with a gaussian significance of $2.6 \sigma$ during a $1362 \; {\rm sec}$
exposure, giving a lower and upper count rate limit of $0.0062 \; {\rm cps}$
and $0.0189 \; {\rm cps}$ respectively (Marshall et al. \cite{EUVE}). For a
plot of the spectral energy distribution see Wolter et al. \cite{wolter}.

\section{Optical Data}
\begin{figure}
\epsfxsize=8.8cm
\epsfysize=5.2cm
\epsfbox{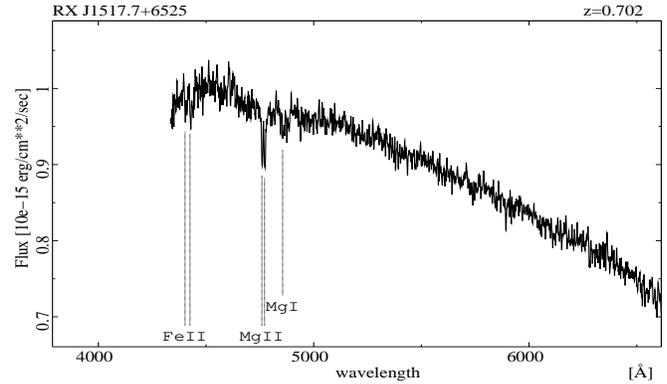}
\caption[]{\label{fig:1998}The spectrum of 1517+656, taken in February 1998 with the 3.5m 
telescope on Calar Alto, Spain using the MOSCA spectrograph. The conditions during the exposure where not photometric, so the flux values can only give a hint to the real flux. The curvature at the blue end below $\sim 4500 {\rm \AA}$ is due to calibration problems. For the doublets see also Fig.~\ref{fig:doublets}} 
\end{figure}

\begin{figure}
\epsfxsize=8.8cm
\epsfysize=5.2cm
\epsfbox{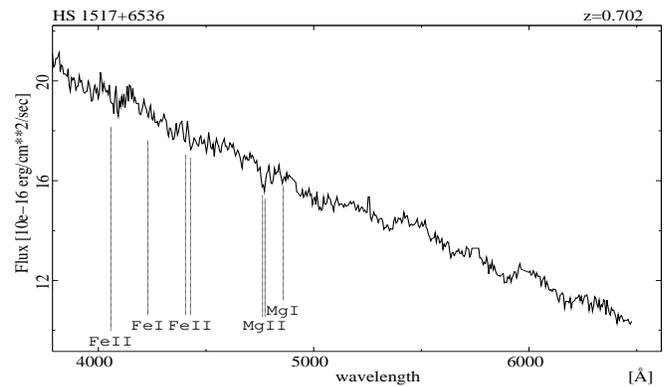}
\caption[]{\label{fig:1993}The spectrum of 1517+656, taken with the 2.2m telescope on Calar Alto in August 1993. Observation conditions were not photometric.}
\end{figure}

\begin{figure}
\epsfxsize=8.8cm
\epsfysize=5.2cm
\epsfbox{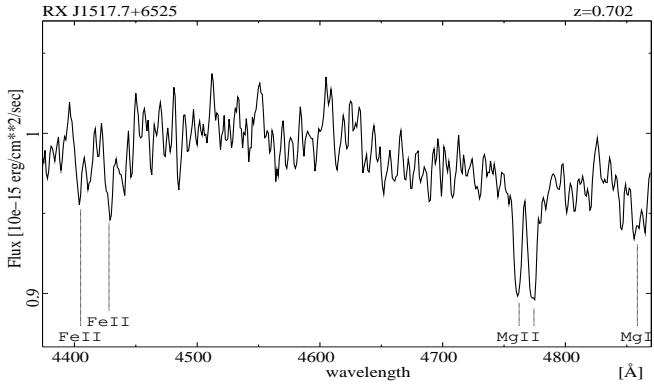}
\caption[]{\label{fig:doublets}Detail of the February 1998 spectrum with the FeII and MgII doublets.} 
\end{figure}

The BL Lac 1517+656 was also included in the Hamburg BL Lac sample selected from the ROSAT All-Sky Survey. This complete sample consists
of 35 objects forming a flux limited sample down to 
$f_{\rm X}(0.5 - 2.0 \; {\rm keV}) = 8 \cdot 10^{-13} \; {\rm erg \; cm^{-2} \; sec^{-1}}$ 
(Bade et al. \cite{bade98}, Beckmann \cite{beckmann}). 
Studying evolutionary effects, we had to
determine the redshifts of the objects in our sample. 
In February 1998 we took
a half hour exposure of 1517+656 with the 3.5m telescope on Calar Alto, Spain,
equipped with MOSCA. Using a grism sensitive in the
$4200 - 6600 \, {\rm \AA}$ range with a resolution of $\sim 3 \, {\rm \AA}$ it was possible to
detect several absorption lines. The spectrum was sky subtracted and flux
calibrated by using the standard star HZ44. Identifying the lines with
iron and magnesium absorption we determined the redshift
of 1517+656 to $z \ge 0.7024 \pm 0.0006$ (see Fig.~\ref{fig:1998}). The part of the spectrum with the FeII and MgII doublet is shown in Fig.~\ref{fig:doublets}.
The BL Lac has also been a target for follow-up observation for the 
Hamburg Quasar Survey (HQS; Hagen et al. \cite{Hagen}) in 1993, because it had 
no published identification then and was independently found by the Quasar selection of the HQS. The $2700\; {\rm sec}$ exposure, taken with the 2.2m 
telescope on Calar Alto and Boller \& Chivens spectrograph, showed a 
power-law like continuum; the significance of the absorption lines in the 
spectrum was not clear due to the moderate resolution of $\simeq 10 \, {\rm \AA}$ (Fig.~\ref{fig:1993}).
Nevertheless the MgII doublet at 4761 
and $4774 {\rm \AA}$ is also detected in the 1993 spectrum, though only marginally resolved \mbox{(see Table 2).}  
The equivalent width of the doublet is comparable in both images ($W_{\rm \AA} = 0.8 / 0.9$ for the 1993/1998 
spectrum respectively).
Also the Fe II absorption doublet at $4403/4228 \, {\rm \AA}$ ($\lambda_{\rm rest} = 2586.6 / 2600.2 \, {\rm \AA}$) and Mg I at 
$4859 \, {\rm \AA}$ ($\lambda_{\rm rest} = 2853.0 \, {\rm \AA}$) is detectable. For a list of the detected lines, see Table 1. 
Comparison with equivalent widths of absorption lines in known elliptical galaxies is difficult because of the underlying non-thermal continuum of the BL Lac jet. But the relative line strengths in the FeII and MgII dublet are comparable to those measured in other absorption systems detected in BL Lac objects (e.g. 0215+015, Blades et al. \cite{blades}).  
\begin{table}
\caption[]{Observed wavelengths and equivalent widths for absorption lines in the February 1998 spectrum}
\begin{tabular}{lllll}
$\lambda_{obs} [\AA]$ & $W_{\lambda} [\AA]$ & $\lambda_{0}$ [\AA] & Ion & Redshift \\  
\hline
4194 & 0.03 & 2463.4 & FeI & 0.7025 \\
- & - & 2484.0 & FeI & not detected\\
4404 & 0.15 & 2586.7 & FeII & 0.7026\\
4429 & 0.17 & 2600.2 & FeII & 0.7033\\
4761 & 0.48 & 2796.4 & MgII & 0.7025\\ 
4774 & 0.52 & 2803.5 & MgII & 0.7028\\
4855 & 0.15 & 2853.0 & MgI & 0.7017\\
4999 & 0.09 & 2937.8 & FeI & 0.7016\\
\end{tabular}
\end{table}
\begin{table}
\caption[]{Observed wavelengths and equivalent widths for absorption lines in the 1993 spectrum}
\begin{tabular}{lllll}
$\lambda_{obs} [\AA]$ & $W_{\lambda} [\AA]$ & $\lambda_{0}$ [\AA] & Ion & Redshift \\  
\hline
4194 & 0.2 & 2463.4 & FeI & 0.7025\\
4231 & 0.1 & 2484.0 & FeI & 0.7033\\
4401 & 0.3 & 2586.7 & FeII & 0.7014\\
4429 & 0.4 & 2600.2 & FeII & 0.7033\\
4761 & 0.4 & 2796.4 & MgII & 0.7025\\ 
4771 & 0.4 & 2803.5 & MgII & 0.7018\\
4855 & 0.15 & 2853.0 & MgI & 0.7017\\
- & - & 2937.8 & FeI & not detected\\
\end{tabular}
\end{table}
Because no emission lines are present and the redshift is measured using absorption lines, the redshift could belong to an absorbing system in the line of sight, as e.g. detected in the absorption line systems in the spectrum of \object{0215+015} (Bergeron \& D'Odorico \cite{bergeron}). A higher redshift would make 1517+656 even more luminous; we will consider this case in the further discussion, though we assume that the absorption is caused by the host galaxy of the BL Lac. 
Assuming a single power law spectrum with $f_{\nu} \propto \nu^{\alpha_{o}}$ the spectral slope in the $4700 - 6600 \, {\rm \AA}$ band can be described by $\alpha_{o} = 0.86 \pm 0.07$. 
The high redshift of this object is even highly plausible, because it was not possible to resolve its host galaxy on HST snap shot exposures (Scarpa et al. \cite{scarpa}).   
The apparent magnitude varies slightly through the different epochs, having reached the faintest value of \mbox{$R=15.9$ mag} and \mbox{$B=16.6$ mag} in February 1999 (direct imaging with Calar Alto 3.5m and MOSCA). These values were derived by comparison with photometric standard stars in the field of view (Villata et al. \cite{villata}).
$H_{0}= 50 \; {\rm km \; sec^{-1} \; Mpc^{-1}}$
and $q_{0} = 0.5$ leads to an absolute optical magnitude of at least $M_{R}= -27.2 \;
{\rm mag}$ and $M_{B} \le -26.4$ (including K-correction).

\section{Mass of 1517+656}

Scarpa et al. (\cite{scarpa}) report the discovery of three arclike structures around 1517+656 in their HST snapshot survey of BL
Lac objects. The radius of this possible fragmented Einstein ring is
2.4~arcsec. If this feature indeed represents an Einstein ring, the
mass of the host galaxy of 1517+656 can easily be estimated. As the
redshift of these background objects is not known, we can only derive a lower
limit for the mass of the lens.

For a spherically symmetric mass distribution (with $\theta$ being the radius of the Einstein ring, $D_{\rm d}$ the angular size distance from the observer to the lens, $D_{\rm s}$ from observer to the source, and $D_{\rm ds}$ the distance from the lens to the source) we get (cf. Schneider et al. \cite{schneider}):
\begin{equation}
M = \theta^2 \frac{D_{\rm d}\,D_{\rm s}}{D_{\rm ds}} \frac{c^2}{4G}
\label{mass}
\end{equation}
Thus the lower limit for the
mass inside the Einstein ring is $M = 1.5 \cdot 10^{12}\,M_{\sun}$ for
Einstein-de~Sitter cosmology and $H_0=50\,\rm
km\,sec^{-1}\,Mpc^{-1}$. For other realistic world models (also
including a positive cosmological constant), this limit is even
higher.

\begin{figure}
\epsfig{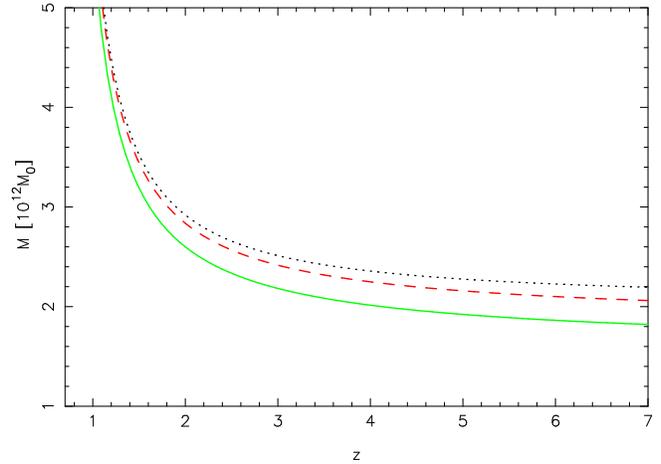}
\caption[]{\label{fig:mass}Mass of the host galaxy of 1517+656 for
  different redshift 
  of the background source. The dotted line is for a
  low-density 
 universe without cosmological constant ($\Omega_{\rm M}=0.3$,
  $\Omega_\Lambda=0$), the dashed one for a flat low-density universe ($\Omega_{\rm M}=0.3$,
  $\Omega_\Lambda=0.7$), and the solid for Einstein-de~Sitter
 cosmology ($\Omega_{\rm M}=1$, $\Omega_\Lambda=0$). We assumed $H_0=50\,\rm km\,sec^{-1}\,Mpc^{-1}$.}
\end{figure}

\begin{figure}
\epsfig{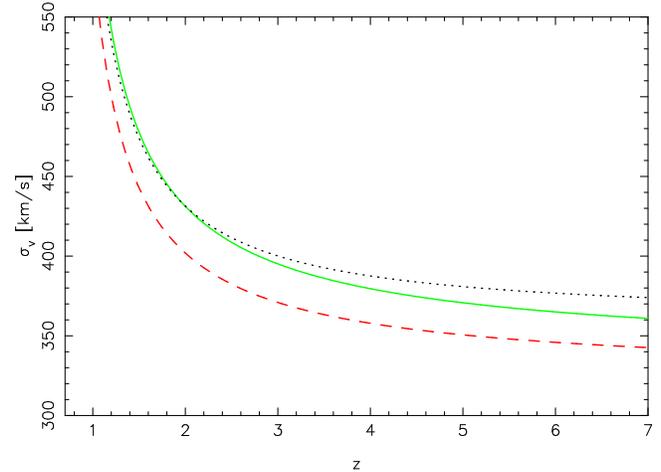}
\caption[]{\label{fig:sigma}Velocity dispersion of the host galaxy of
  1517+656 for the same cosmological models as in Fig.~\ref{fig:mass}.}
\end{figure}

Assuming an isothermal sphere for the lens, the velocity dispersion
in the restframe can be calculated by
\begin{equation}
\sigma_v^2 = \frac{\theta}{4\pi} \frac{D_{\rm s}}{D_{\rm ds}} c^2
\end{equation}
Independent of $H_0$ we get a value of at least $330\,\rm km\,sec^{-1}$ for
Einstein-de~Sitter cosmology, and slightly less 
($320\,\rm km\,sec^{-1}$) for a flat low-density
universe ($\Omega_{\rm M}=0.3$,  $\Omega_\Lambda=0.7$).  Other models
again lead to
even higher values.  The true values of the mass and velocity
dispersion might be much higher if the redshift of the source is
significantly below $z \approx 2$. Figures~\ref{fig:mass} and
\ref{fig:sigma} show the mass and velocity dispersion as a function of
the source redshift.

If the observed absorption is caused by a foreground object and the redshift
of 1517 is higher than 0.7, the mass and velocity dispersion of
the host galaxy have to be even higher.

More detailled modelling of this system will be possible when the
redshift of the background object is measured. If the arcs are caused
by galaxies at different redshift, the mass distribution in the outer
parts of the host galaxy of 1517+656 can be determined which will provide
very important data for the understanding of galaxy halos.
High resolution and high S/N direct images may allow to use more
realistic models than symmetrical mass distributions by providing
further constraints.

\section{Discussion}
The BL Lac 1517+656  with $M_{R} \le -27.2 \;{\rm mag}$ and $M_{B} \le -26.4$ is the most luminous BL Lac object in the optical band.
Padovani \& Giommi (\cite{padovani}) presented in their catalogue of 233
known BL Lacertae objects an even brighter candidate than \object{1ES1517+656}: \object{PKS 0215+015} (redshift
$z= 1.715$, $V = 15.4 \; {\rm mag}$, V\'eron-Cetty \& Veron
\cite{veron93}). This radio source has been identified by Bolton \& Wall
(\cite{bolton}) as an $18.5 \; {\rm mag}$ QSO. The object has been mainly in a
bright phase starting from 1978, and became faint again since mid-1983 (Blades et
al. \cite{blades}). Its brightness is now $V = 18.8 \; {\rm mag}$ ($M_{V} = -26.2 \; {\rm mag}$; Kirhakos et
al. \cite{kirhakos}, V\'eron-Cetty \& Veron \cite{veron98}).\\
Also the X-ray properties of 1517+656 are extreme: with an X-ray flux of 
\mbox{$f_{\rm X} = 2.89 \; \cdot
10^{-11} \; {\rm erg \; cm^{-2} \; sec^{-1}}$} in the ROSAT PSPC band we have a
luminosity of $L_{X}= 7.9 \; \cdot 10^{46} \; {\rm erg \; sec^{-1}}$ which is 
a monochromatic luminosity at $2 \; {\rm keV}$ of $L_{X}= 4.6 \; \cdot 10^{21} \; {\rm W \; Hz^{-1}}$. 
The radio flux of $37.7 \rm{mJy}$ at $1.4 \; {\rm GHz}$ leads to $L_{R} = 
1.02 \; \cdot 10^{26} \; {\rm W \; Hz^{-1}}$.  
Thus 1517+656 is up to now one of the most luminous known BL Lac in X-rays, radio and optical band, also compared to newest results from HST observations (Falomo et al. \cite{falomo}). They give detailed analysis for more than 50 BL Lac objects with redshift $z < 0.5$, showing none of them having an absolute magnitude $M_{R} < -26$. Compared to the 22 BL Lac in the complete EMSS sample (Morris et al. \cite{morris}), 1517+656 is even more luminous in the radio, optical and X-ray band than all of those high frequency peaked BL Lac objects (HBL).
Finding an HBL, like 1517+656 with 
$\nu_{\rm peak} = 4.0 \cdot 10^{16} \; {\rm Hz}$ (Wolter et al. \cite{wolter}), of such brightness is even more surprising, because the HBL are usually
thought to be less luminous than the low frequency peaked ones (e.g. Fossati et al. \cite{fossati}, Perlman \&
Stocke \cite{perlman}, Januzzi et al. \cite{januzzi}).
In comparison to the SED for different types of Blazars, as shown in Fossati et al. (\cite{fossati}), 1517+656 shows a remarkable behaviour. The radio-properties are similar to an HBL ($\log(\nu L_{4.85 \;{\rm GHz}}) = 42.7$), in the V-Band ($\log(\nu L_{5500 \; \AA}) = 46.1$) and in the X-rays ($\log(\nu L_{1 \; {\rm keV}}) = 46.4$) between bright LBL and faint FSRQ objects.  

On the other hand it is not surprising to find one of the most luminous BL Lac objects in a very massive galaxy with $M > 2 \cdot 10^{12} M_{\sun}$. This mass is a lower limit, as long as the redshift of 1517+656 could be larger than $z=0.702$, and is depending on the cosmological model and on the redshift of the lensed object (see Fig. 4). 

\begin{acknowledgements}We would like to thank H.-J. Hagen for developing the
optical reduction software and for taking the 1993 spectrum of \object{HS 1517+656}. Thanks to Anna Wolter and the other colleagues from the {\it Osservatorio Astronomico di Brera} for fruitful discussion.
This work has received partial financial support from the Deutsche Akademische Austauschdienst.
\end{acknowledgements}

\end{document}